\documentclass[twocolumn]{aastex631}

\let\tablenum\relax
\usepackage{dirtytalk}
\usepackage{graphicx}
\usepackage{float}
\usepackage{enumitem}
\usepackage{CJK}
\usepackage{siunitx}
\usepackage{siunitx}
\usepackage{multirow}

\begin{document}
\begin{CJK*}{UTF8}{gbsn}

\title{On the roles of stellar rotation and binarity in NGC 2423's main-sequence turnoff region}

\correspondingauthor{Chengyuan Li}
\email{lichengy5@mail.sysu.edu.cn}

\author[0009-0001-7230-3348]{Yutian Bu (卜玉田)}
\affiliation{School of Physics and Astronomy, Sun Yat-sen University, Zhuhai, 519082, Peopleʼs Republic of China}
\affiliation{CSST Science Center for the Guangdong-Hong Kong-Macau Greater Bay Area, Zhuhai, 519082, China}

\author[0000-0001-9131-6956]{Chenyu He (贺辰昱)}
\affiliation{School of Physics and Astronomy, Sun Yat-sen University, Zhuhai, 519082, Peopleʼs Republic of China}
\affiliation{CSST Science Center for the Guangdong-Hong Kong-Macau Greater Bay Area, Zhuhai, 519082, China}

\author[0000-0003-3471-9489]{Li Wang (王莉)}
\affiliation{School of Physics and Astronomy, Sun Yat-sen University, Zhuhai, 519082, Peopleʼs Republic of China}
\affiliation{CSST Science Center for the Guangdong-Hong Kong-Macau Greater Bay Area, Zhuhai, 519082, China}

\author{Jiamao Lin (林佳茂)}
\affiliation{School of Physics and Astronomy, Sun Yat-sen University, Zhuhai, 519082, Peopleʼs Republic of China}
\affiliation{CSST Science Center for the Guangdong-Hong Kong-Macau Greater Bay Area, Zhuhai, 519082, China}

\author[0000-0002-3084-5157]{Chengyuan Li (李程远)}
\affiliation{School of Physics and Astronomy, Sun Yat-sen University, Zhuhai, 519082, Peopleʼs Republic of China}
\affiliation{CSST Science Center for the Guangdong-Hong Kong-Macau Greater Bay Area, Zhuhai, 519082, China}



\begin{abstract}
Research has shown that many young and intermediate-age clusters (younger than $\sim$2 Gyr) have extended main sequences and main-sequence turnoffs (eMSTOs), which cannot be adequately described by a single isochrone. The reason for the extended main sequences is now known, with the most probable cause being the fast rotation of stars. However, a significant fraction of slowly rotating stars form a younger stellar population than their fast-rotating counterparts, leading to speculation that they have undergone thorough rotational mixing processes internally. One speculation is that a considerable number of slowly rotating stars reside in close binary systems, where tidal forces from companion stars are the cause of their rotational deceleration. In this work, we report a relatively old open star cluster in the Milky Way, NGC 2423 ($\sim$1 Gyrs old), which exhibits an apparent eMSTO. As anticipated, many characteristics of NGC 2423 indicate that its eMSTO is driven by stellar rotations. Our calculations indicate that if slowly rotating stars commonly have a close companion star, they should exhibit significant differences in radial velocities observationally, and binary systems that can be tidally locked within the age of NGC 2423 should have a mass ratio close to 1. However, none of these predictions align with our observations. Interestingly, among the only two equal-mass binary systems in the observed region for which spectroscopic data could be obtained, we discovered that one of them is a tidally locked binary system. This further suggests the validity of our numerical simulation results.
\end{abstract}




\section{Introduction}
It has been two decades since the phenomenon of extended main-sequence turnoffs (eMSTOs) was first discovered in the Magellanic Cloud clusters \citep{2003AJ....125..770B}. This phenomenon was soon proven to be quite common in young star clusters in both the Magellanic Clouds and the Milky Way \citep{2009A&A...497..755M,2018ApJ...869..139C}. The origin of eMSTOs was a subject of heated debate \citep[e.g., ][]{2009MNRAS.398L..11B,2011MNRAS.412L.103G,2014Natur.516..367L,2015MNRAS.450.1693G,2015ApJ...807...25B,2017ApJ...846...22G,2022ApJ...934..105L}, until recently when it was widely realized as a physical phenomenon driven by stellar rapid rotations, specifically the rotation related gravity darkening \citep[e.g.,][]{1968ApJ...151..203F}, mixing \citep[e.g.,][]{2013ApJ...776..112Y}, and the formation of self-extinguishing stellar disks due to rotation\citep[e.g.,][]{Milone2023A&A...672A.161M,Antona2023MNRAS.521.4462D}. For an overview of the historical development, relevant hypotheses, debates, and scientific discussions on various models in this dynamic field, one can consult the review article by \cite{LI2024}

Considering only the gravitational darkening effect resulting from stellar rotation, we can effectively account for the observed eMSTO phenomenon. However, the blue section of the eMSTOs, representing slowly rotating stars \citep{2018ApJ...863L..33M,2018AJ....156..116M, 2019ApJ...883..182S,2023MNRAS.518.1505K}, appears younger than anticipated, presenting new challenges to our understanding of how stellar rotation impacts internal stellar structure. The calculation of \cite{D'Antona_2017NatAs...1E.186D} suggest that to generate a blue stellar population that appears younger than the bulk population stars (while rotating slowly), it is necessary to assume that these stars have undergone significant rotational mixing, indicating they were previously fast rotators. However, the reasons for these stars' deceleration remains uncertain. They speculate that many of these blue stars might have a close companion, with tidal interactions from these companions potentially driving the deceleration in rotation of the primary stars.

It is essential to examine the \cite{D'Antona_2017NatAs...1E.186D}'s hypothesis, as binary star systems that undergo tidal deceleration during the main-sequence (MS) stage must have very different evolutionary paths compared to single stars. These binary systems could cause significant deviations in the stellar population of the cluster from the single-stellar population scenario. However, such a testing available for this hypothesis is rare. Only two clusters have been utilized for directly validating this hypothesis. \cite{2021MNRAS.508.2302K} studied a Large Magellanic Cloud (LMC) cluster, NGC 1850, a young star cluster {$\sim$100 Myr-old} with a bifurcated MS. By employing time-domain spectroscopic measurements, they found no significant differences in the binary fraction between the blue and red MSs (bMS and rMS). They concluded that binaritity is not the main factor contributing to the significant differences in rotations between the bMS and rMS. \cite{hechenyu_2023MNRAS.525.5880H} studied the open star cluster NGC 2422 ($\sim$90 Myr-old) in the Milky Way, which exhibits a similar split MS with stars showing a wide distribution of rotation rates. Using time-domain spectroscopy, they also rule out the possibility of a significant population of tidally locked binaries in this cluster. By integrating numerical $N$-body simulations with stellar evolution models, \cite{wangli_2023ApJ...949...53W} have examined the potential presence of tidally locked binaries among bMS stars. The testbed object of their study is the LMC young massive cluster NGC 1856 ($\sim$300 Myr-old), and they yield a negative result for this cluster.

In addition to indirectly ruling out the dominance of tidally locked binary systems among bMS stars in NGC 1856, \cite{wangli_2023ApJ...949...53W}'s calculations also revealed an interesting fact that only binary systems with a mass ratio close to unity ($q\sim1$, or equal-mass binaries) in young star clusters can reach tidal locking within such a short timeframe. If their findings hold true, it suggests the possibility of locating these ``twins'' near the equal-mass binary sequence in star clusters, which is of significant importance for understanding the formation of ultracompact binaries (i.e., double white dwarfs).

Clearly, so far the significant role of tidal binaries in modulating stellar rotations is only ruled out in a very limited number of clusters, all of which happen to be quite young. Since tidal locking involves a long-term modulation of binary orbital periods, it remains unclear whether tidal locking could play a more important role in the distribution of stellar rotation within older star clusters ($\sim$1--2 Gyr-old). Additionally, given that numerical simulations suggest the presence of some equal-mass tidally locked binaries in young star clusters, theoretically, these ``twins'' should be more prevalent in older star clusters (assuming tidal locking persists once established). Therefore, a direct observational search for such ``twins'' is of crucial importance in validating the numerical simiulation.

In this study, we report an old Milky Way open star cluster, NGC 2423 ($\sim$1 Gyr-old), which exhibits a clear eMSTO. As a test case, our findings are as follows: Firstly, we confirm that the eMSTO is due to different stellar rotations, consistent with previous studies. Secondly, through statistical analysis, we rule out the possibility that the slow-rotating stars (referred to as bMS stars) in the cluster are predominantly tidally locked binaries. Additionally, we conducted $N$-body numerical simulations combined with stellar evolution models for this cluster. The simulations suggest that for a 1 Gyr-old cluster, only those binaries with very close to equal masses are likely to be tidally locked (also), at least under non-extreme initial conditions (e.g., where the initial semi-major axis distribution of binaries is not heavily biased towards the close side). Lastly, and most notably, we indeed discovered a ``twins'' in this cluster at a position close to the equal-mass binary loci. Although only two such equal-mass binaries with accessible spectra were found in our field of view, this discovery may imply the accuracy of our models.

This paper is organized as follows. In section \ref{sec:methods}, we present the membership determination, isochrone fitting, the stellar projected rotational velocity calculation and $N$-body simulation. In section \ref{sec:results}, we present our main result about the origin of eMSTO regarding age spread and stellar rotation and the synthetic and observational tidally-locked binaries. Our discussion and conclusion are presented in section \ref{sec: discussion} and section \ref{sec: conclusion} respectively.

\section{Methods} \label{sec:methods}

\begin{figure*}[t]
    \centering
    \includegraphics[width=\textwidth]{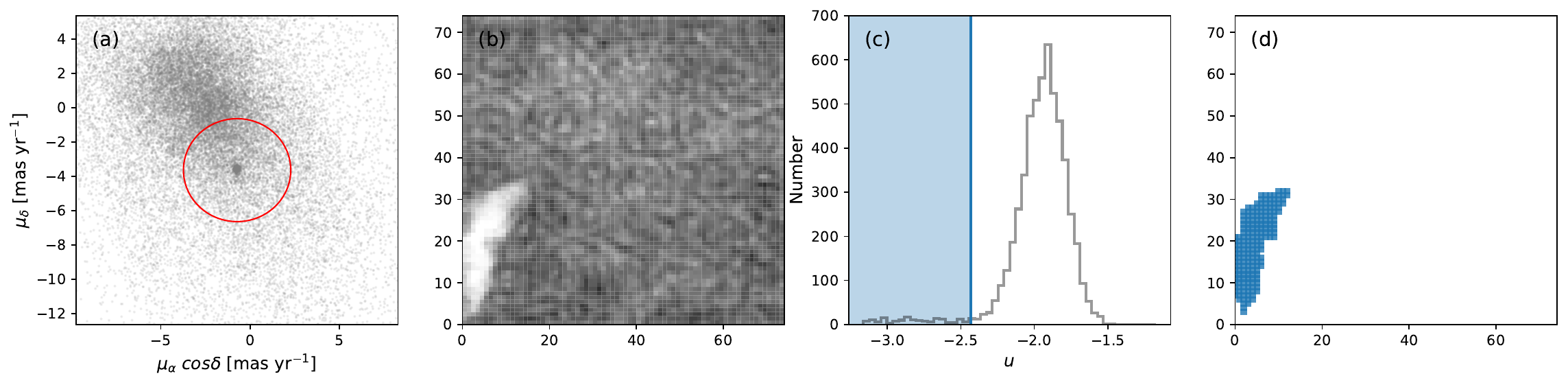}
    \caption{(a) PM vector plot for the region around NGC 2423. The red circle, with a radius of 3 mas yr$^{-1}$, marks the PM selection range. (b) 2D neural network resulting from SOM, where the $u$ value between neighboring neurons is represented by grayscale. (c) Distribution histogram of $u$ values. The blue solid line is the selection of $u$ that produces a 5\% contamination rate among the identified member candidates, for the blue patch (d) in the 2D neural network.
    }
    \label{f1}
\end{figure*}

\subsection{Membership Determination} \label{sec:membership}

The membership determination of NGC 2423 is based on Gaia early data release 3 (EDR3) \citep{2021A&A...649A...1G}. The distance to NGC 2423 is taken to be 924 pc, and the equatorial coordinates of its center are R.A. = 114.3$^{\circ}$, Decl. = $-$13.8$^{\circ}$ \citep{10.1093/mnras/stab770}. For the selection of member candidates, we initially select a region with a radius of 100 pc centered at Galactocentric coordinates (X, Y, Z) = ($-609.2, -737.9, 59.7$) pc for NGC 2423. These coordinates are derived from R.A., Decl., and parallax data utilizing the Python {\tt\string Astropy} package \citep{astropy2013, Price-Whelan_2018} for transformation. A subsequent selection criterion is applied based on proper motion (PM). Figure \ref{f1}(a) visually represents the PM distribution for the regions around NGC 2423. Subsequently, stars located within a radius of 3 mas yr$^{-1}$ from the PM center of ($\mu_\alpha$cos$\delta$, $\mu_\delta$) = (114.29, $-$13.83) mas yr$^{-1}$ \citep{10.1093/mnras/stab770} are chosen. 

In this study, we implement a series of measures to exclude possible artifacts. We adopt the quality criteria proposed in Appendix C of \citet{lindegren_gaia_2018} to select stars with parallaxes and photometric measurements less than 10 percent uncertainty. The final selected sample consists of 5,531 stars.

The unsupervised machine learning method, {\tt\string StarGO}\footnote{\url{https://github.com/zyuan-astro/StarGO-OC}} \citep{Yuan_2018}, is employed to identify cluster members of NGC 2423. Notably, {\tt\string StarGO} has exhibited its success in identifying memberships within both open clusters \citep{Tang_2019, Pang_2020,Pang_2021a,Pang_2021b} as well as stellar streams \citep{Yuan_2020a,Yuan_2020b}. The foundation of this algorithm rests upon the Self-Organizing Map (SOM), which facilitates the projection of high-dimensional data onto a two-dimensional neural network, all while preserving the intrinsic topological structures inherent within the dataset. Consequently, stars that cluster together in the high-dimensional space are associated with neurons grouped on the two-dimensional map.

We apply the {\tt\string StarGO} to map a 5D dataset (X, Y, Z, $\mu_{\alpha} \rm{cos}\delta$, $\mu_{\delta}$) onto a 2D neural network. In brief, we initially generate a 75 $\times$ 75 network to scale the number of neurons to the number of stars in our input sample, wherein each neuron possessed a weight vector of equivalent dimensionality to the input vector. Sequentially, stars from our sample were fed, one at a time, to all 5,625 neurons. Each neuron underwent an adjustment of its weight vector to approximate the input vector associated with a specific star. A single iteration was deemed complete upon the neurons being exposed to all stars in our input sample once. The entire learning cycle was repeated 400 times until the weight vectors converged. We illustrate the trained neural network in Figure \ref{f1}(b). The magnitude of the disparity in weight vectors between neighboring neurons, denoted as `$u$’, diminishes when the 5D weight vectors of adjoining neurons exhibit similarity. This correspondence signifies that stars linked with these neurons exhibit congruence in spatial and kinematic attributes. Neurons with small $u$ values will congregate on the 2D SOM map, forming the distinct lighter-shaded patch in Figure \ref{f1}(b). Neurons situated within the patch exhibit diminished values of $u$, while those outside the patch manifest larger values of $u$. The $u$ values associated with neurons inside the patch contribute to the elongated tail toward smaller values in the $u$-histogram of Figure  \ref{f1}(c). The patch (blue color in Figure \ref{f1}(d)) was further delineated by the cut on the extended distribution of $u$.

\begin{figure}
    \centering
    \includegraphics[width=0.45\textwidth]{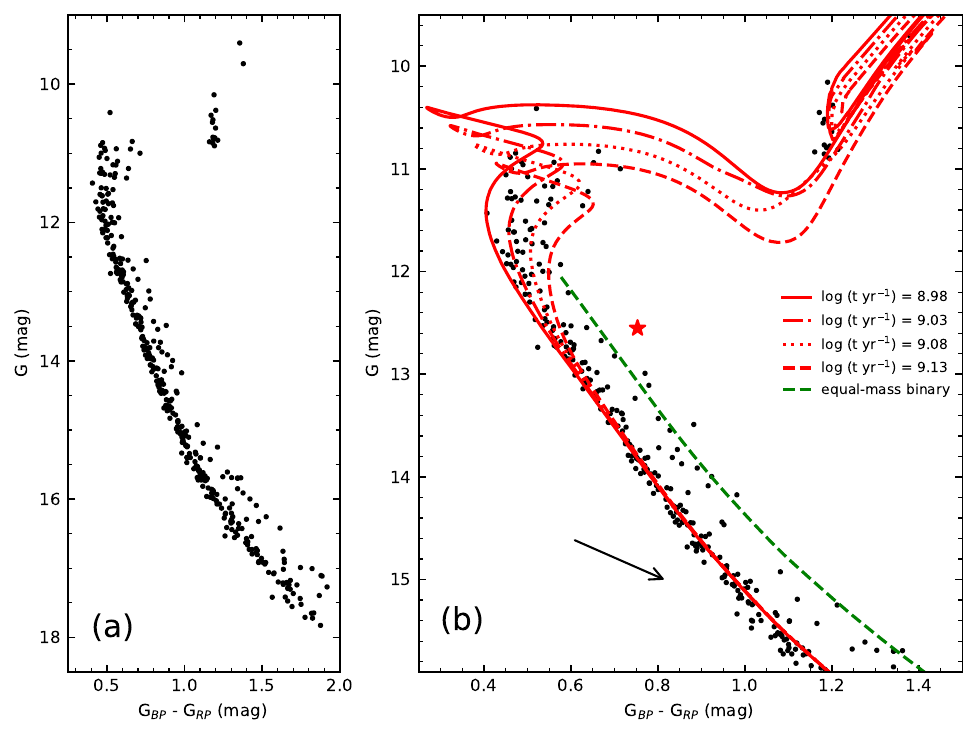}
    \caption{(a), The CMD of NGC 2423 for a comprehensive view. (b), The CMD showing isochrone fitting for various ages. The black arrow indicates the reddening vector, and the red pentagram represents the identified tidally locked binary.
    }
    \label{fig:members}
\end{figure}

A group of member stars in NGC 2423 corresponds to a local minimum within the $u$ values, associating with neurons that possess a $u$ value below a specific threshold, which is determined by an approximate 5\% contamination rate from field stars. The field star contamination rate is computed employing stars from the smooth Galactic disk population with the Gaia DR2 mock catalog  \citep{Rybizki_2020}. Following the application of the identical selection criteria described in Section \ref{sec:membership} to the mock stars, the mock stars that remain associated with the neurons in a manner akin to the cluster members are recognized as field star contaminants. A total of 396 stars have been selected as member candidates of NGC 2423. The color-magnitude diagram (CMD) of those candidates clearly shows an eMSTO at G$\sim$11.5 mag as shown in Figure \ref{fig:members}(a).

\subsection{Isochrone Fitting} \label{sec:isochrone}
We use the isochrones from  MESA Isochrones \& Stellar Tracks (MIST, version 1.2) \citep{2016ApJ...823..102C} with initial rotation velocity of 0.0 critical velocity to fit the CMD of the cluster, as presented in Figure \ref{fig:members}(b). The morphology of the eMSTO can not be fitted by a single isochrone. The best-fitting isochrone to the blue edge of the MSTO region shows that the cluster has an age of $\log (t\ yr^{-1}) = 8.98$ ($\sim$1 Gyr), metallicity of [Fe/H] = 0.18 dex, and extinction $A_{V}$ = 0.15 mag, at a distance of 860 pc (corresponding to $(m-M)_0\sim$9.7 mag). Our results are in agreement with \cite{10.1093/mnras/stab770} which gave an age of  $\log (t\ yr^{-1}) = 9.037$, D = 924 pc and [Fe/H] = 0.122 dex for NGC 2423. 

While differential reddening could potentially affect the morphology of the MSTO region, the main-sequence (MS) and the red clump of NGC 2423 appears relatively compact in comparison. The influence of differential reddening on NGC 2423 is minimal. In Figure \ref{fig:members}(b), the reddening vector ($\Delta{A}_{V}$ = 0.5 mag) is depicted. Binaries might also contaminate the MSTO region, but the effect is weak \citep{2009A&A...497..755M}. The equal-mass binary sequence of NGC 2423 is generated by shifting the best-fitting isochrone by $-0.75$ mag in G-band magnitude, as the green line shown in Figure \ref{fig:members}(b). The extend of MSTO is too large to be caused by binaries.

\subsection{Projected Rotational Velocities} \label{sec:vsini}

We derived the spectra from European Southern Observatory archive (PI: Zaworska Paulina, ID: 098.C-0217) to calculate the projected rotational velocities ($v\sin i$) of 27 stars in the MS and MSTO region. The wavelengths of the spectra are between 644.2 nm and 682.2 nm with a resolution of 17,000 and a signal to noise ratio (SNR) of greater than 50. The observations were taken through the GIRAFFE spectrograph on the Very Large Telescope in 2017. 

The absorption line of observational spectra we choose to calculate $v\sin i$ is the Fe \uppercase\expandafter{\romannumeral1} absorption lines because it is prominent in the spectra and nearly unaffected by atmospheric absorption line. The Fe \uppercase\expandafter{\romannumeral1} absorption line ($\sim$6678\AA) is fitted by synthetic spectra with different parameters, generated from the POLLUX database \citep{2010A&A...516A..13P}.   The effective temperature $T_\mathrm{eff}$ of the symthetic spectra ranges from 5,800 K to 8,500 K in steps of 100 K. The surface gravity $\log g$ ranges from 3.5 dex to 5.0 dex in steps of 0.1 dex. [Fe/H] ranges from $-0.5$ dex to 0.5 dex in steps of 0.5 dex. 

\begin{figure}
    \centering
    \includegraphics[width=0.4\textwidth]{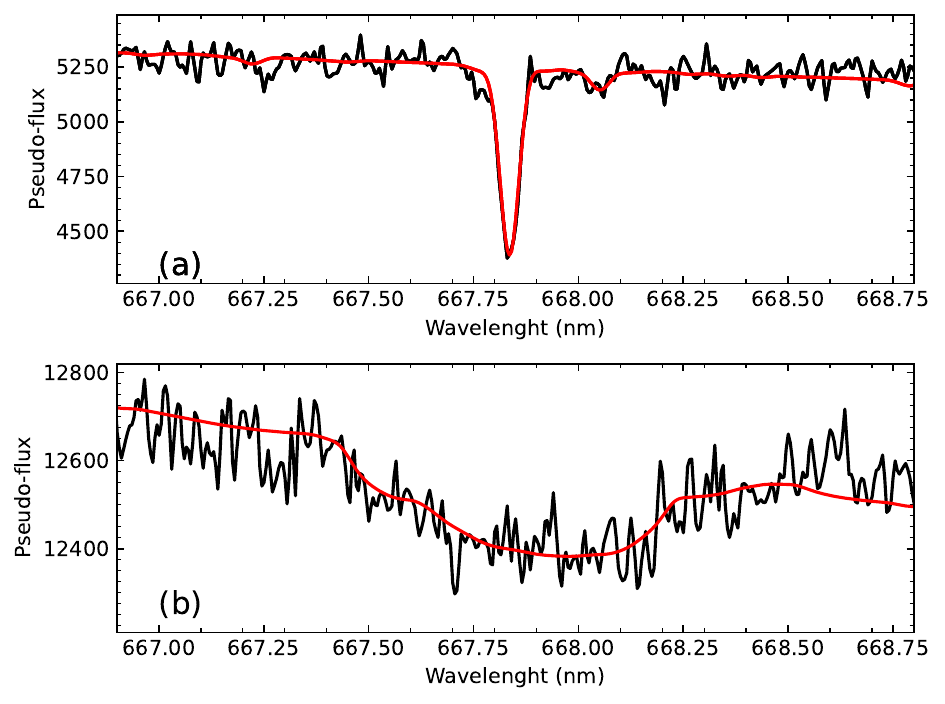}
    \caption{This figure illustrates how we obtained the projected stellar rotation velocities through fitting. Panels (a) and (b) show the spectra and model fits (red curves) for a slowly rotating star and a rapidly rotating star, respectively.
    }
    \label{fig:vsini_spectrum}
\end{figure}

To get stellar projected rotational velocities, considering that the mean radial velocity (RV) of NGC 2423 is about \SI{20}{\kilo\meter\per\second}\citep{10.1093/mnras/stab770}, we apply Doppler shift corresponding to radial velocities ranging from \SI{15}{\kilo\meter\per\second} to \SI{25}{\kilo\meter\per\second} in steps of \SI{1}{\kilo\meter\per\second}, instrumental broadening corresponding to a resolution of 17,000 and stellar rotating broadening to the synthetic spectra using PyAstronomy\citep{2019ascl.soft06010C}. The $v\sin i$ are firstly chosen from \SI{10}{\kilo\meter\per\second} to \SI{350}{\kilo\meter\per\second}, in steps of \SI{5}{\kilo\meter\per\second}. Afterwards, we use Astrolib PySynphot \citep{2013ascl.soft03023S} to calculate the flux of the synthetic spectra corresponding to the observed wavelengths. We use $\chi ^2$-minimization method to determine the best-fitting model to each observed spectrum. 
After obtaining the best-fitting $v\sin i$ from the first run of fitting, we enhance the calculation precision by incrementally reducing the step size by half of the previous calculation within the range of the best-fitting $v\sin i$ plus and minus the step size of the previous run. Through 13 iterations, the step size is smaller than 0.001 \SI{}{\kilo\meter\per\second}. The uncertainty of the $v\sin i$ is determined by the difference between the lower and upper $v\sin i$ values with $\chi^2$ being equal to $\chi^2_{\rm min}$+1. The results are listed in Table \ref{tab:full_table}. An illustrative fitting result is depicted in Figure \ref{fig:vsini_spectrum}.

\begin{deluxetable*}{ccccccc} 
\caption{Information of NGC 2423 member stars with $v\sin i$ measurement}
\label{tab:full_table}

\tablehead{
\colhead{Gaia EDR3 ID} & \colhead{RA [deg]} & \colhead{Dec [deg]} & \colhead{G [mag]} & \colhead{G$_{BP}-$G$_{RP}$ [mag]} & \colhead{$v\sin i$ [\SI{}{\kilo\meter\per\second}]} & \colhead{$e_{v\sin i}$ [\SI{}{\kilo\meter\per\second}]}
}

\startdata
3030268717760216064	& 114.281184393985 	& -13.851504744727            & 11.614471 	& 0.468177 	    & 11.08 	&	1.41 	\\
3030260269569121152	&	114.201687008885 	&	-13.978059142776 	&	14.123535 	&	0.800491 	&	13.47 	&	2.00 	\\
3030280954131465088	&	114.260932663378 	&	-13.800998393915 	&	13.677393 	&	0.725358 	&	14.82 	&	2.48 	\\
3030280679253552000	&	114.280569756578 	&	-13.805425078016 	&	12.850809 	&	0.594273 	&	18.28 	&	1.51 	\\
3030273738586219904	&	114.092610282778 	&	-13.883815944722 	&	13.818087 	&	0.770230 	&	18.89 	&	1.85 	\\
3030268859503564032	&	114.302316714733 	&	-13.831953142603 	&	13.648598 	&	0.730248 	&	19.12 	&	1.84 	\\
3030261197282012800	&	114.300291275690 	&	-13.975733730469 	&	12.656590 	&	0.562656 	&	20.86 	&	1.27 	\\
3030257280271822976	&	114.374577402505 	&	-13.997312960615 	&	12.827416 	&	0.586217 	&	28.29 	&	2.25 	\\
3030274769378635648	&	114.185018541137 	&	-13.854332961347 	&	11.706479 	&	0.474842 	&	31.78 	&	1.17 	\\
3030288581993093760	&	114.224627386337 	&	-13.609784677808 	&	12.526540 	&	0.535455 	&	69.11 	&	2.35 	\\
3030280988491186432	&	114.302667098947 	&	-13.806446566910 	&	13.098605 	&	0.632124 	&	70.25 	&	7.72 	\\
3030280434430999424	&	114.250918309993 	&	-13.850600034101 	&	11.166129 	&	0.539528 	&	74.85 	&	4.02 	\\
3030282191082013824	&	114.259802949306 	&	-13.757962509459 	&	10.927766 	&	0.485625 	&	78.44 	&	2.31 	\\
3030269443619073920	&	114.429966518293 	&	-13.856141511809 	&	11.220549 	&	0.461205 	&	85.53 	&	4.61 	\\
3030257997521353472	&	114.355888691514 	&	-13.986392396261 	&	12.687275 	&	0.598741 	&	93.40 	&	6.53 	\\
3030268722064641024	&	114.276277437470 	&	-13.856193038168 	&	12.733048 	&	0.589325 	&	99.66 	&	12.24 	\\
3030261609598838400	&	114.302922265973 	&	-13.924738774482 	&	10.410906 	&	0.520599 	&	117.07 	&	4.29 	\\
3030269168741204096	&	114.321846570120 	&	-13.838246164879 	&	11.754764 	&	0.542961 	&	121.44 	&	7.42 	\\
3030274288342312960	&	114.180680570758 	&	-13.873002570704 	&	11.580619 	&	0.510057 	&	125.14 	&	6.83 	\\
3030257967466601344	&	114.332125025510 	&	-13.989737186005 	&	12.594372 	&	0.588397 	&	138.32 	&	9.19 	\\
3030267244595909760	&	114.355346956374 	&	-13.928238533216 	&	10.938271 	&	0.652101 	&	150.96 	&	3.90 	\\
3030280885411983104	&	114.277002140949 	&	-13.804608900770 	&	11.115024 	&	0.568714 	&	166.75 	&	6.74 	\\
3030274735018908032	&	114.180213080946 	&	-13.866505517940 	&	12.115068 	&	0.503320 	&	166.86 	&	9.83 	\\
3030264289658416256	&	114.362371232297 	&	-13.939672544778 	&	11.524990 	&	0.539303 	&	167.34 	&	7.28 	\\
3030260544447025920	&	114.181246134077 	&	-13.962408661998 	&	11.347738 	&	0.548876 	&	177.89 	&	9.44 	\\
3030271436483827840	&	114.403858610401 	&	-13.745216947146 	&	11.728625 	&	0.510724 	&	279.79 	&	8.52 	\\
3030269374902128384	&	114.341598953068 	&	-13.806532012783 	&	11.996540 	&	0.535441 	&	306.90 	&	10.96 	\\
\enddata

\tablecomments{The information of 27 stars with $v\sin i$ measurement. G is the Gaia G-band magnitude. G$_{BP}$ is the magnitude of Gaia blue pass band and G$_{RP}$ is the magnitude of Gaia red pass band. $e_{v\sin i}$ is the uncertainty of $v\sin i$}.
\end{deluxetable*}

\subsection{N-body Numerical Simulation} \label{sec:simulation}
To investigate the properties of tidally-locked binaries candidate in NGC 2423, we generated a NGC 2423-like synthetic cluster, using the N-body code {\tt\string PeTar}\footnote{\url{https://github.com/lwang-astro/PeTar}} \citep{wanglong_2020MNRAS.497..536W}. {\tt\string PeTar} is designed to model collisional stellar systems where multiplicity and close encounters are crucial for dynamical evolution. In the adopted model, the cluster evolution is governed by stellar evolution, stellar interaction and the external Galactic potential.

In simulation, the initial condition file is generated from the cluster producing tool {\tt\string McLuster} \citep{mcluster_2011MNRAS.417.2300K,mcluster_2019MNRAS.484.1843W}. The initial total mass is set to \SI{2000}{M_\odot} and metallicity to 0.03. Those parameters are empirically determined by the present cluster mass and number density profile of NGC 2423. The half mass radius is \SI{0.27}{pc}, derived from \cite{marks_2012A&A...543A...8M} relation. The Plummer density profile \citep{Aarseth_1974A&A....37..183A} is applied to generate positions and velocities of stars. The initial masses of stars were drawn from a Kroupa-like initial mass function \citep{kroupa_2001MNRAS.322..231K} spanning a mass range of 0.8--150 \SI{}{M_\odot}. The initial binary fraction is set to 100\%. The simulation continued until the synthetic cluster reached an age of \SI{960}{Myr}, similar with the age of NGC 2423 as determined by the best-fitting isochrone, producing snapshots of both single and binary data. We have confirmed that the NGC 2423-like cluster has similar structural parameters to the actual observation. The PARSEC Bolometric Correction \citep{YBC_2019A&A...632A.105C} is utilized to convert the luminosity and temperature into Gaia EDR3 filters.

\section{Results} \label{sec:results}
\subsection{The impact of different ages and rotation} \label{sec:rv}
\begin{figure}
    \centering
    \includegraphics[width=0.45\textwidth]{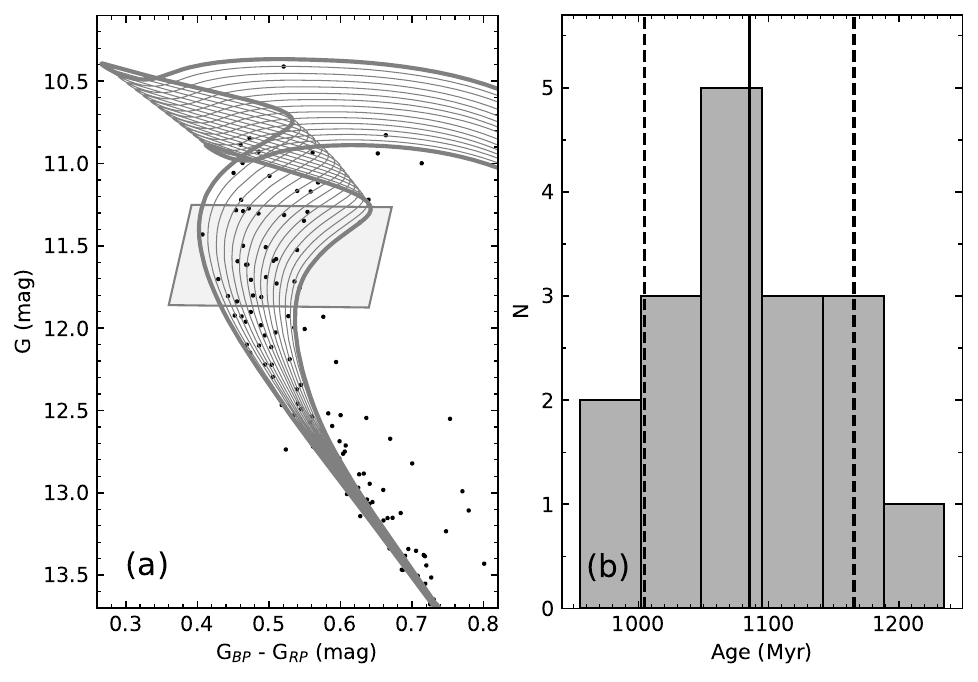}
    \caption{(a) Overlaid on the CMD of NGC 2423 are grids of MIST isochrones (grey curves) with age step of 25 Myr, used for estimating any potential age variation within NGC 2423. (b) The histogram illustrates the distribution of ages among the calculated stars. The solid line denotes the median age (1085 million years), while the two dashed lines represent the median $\pm{\sigma}$ (80.62 Myr).
    }
    \label{fig:fwhm1}
\end{figure}

Under the assumption of prolonged cluster formation, the extend of the MSTO represents an age spread. To quantify the age spread in the MSTO region, by over-plotting the isochrones in steps of 5 Myr, we estimated the ages of 27 stars located in the grey parallelogram area as shown in Figure \ref{fig:fwhm1}(a). The uncertainty of the age is 5 Myr. For clarity, we only show the isochrones in steps of 25 Myr in Figure \ref{fig:fwhm1}(a). Figure \ref{fig:fwhm1}(b) shows the distribution of the stellar ages based on our estimation. The photometric errors of NGC 2423 from Gaia EDR3 are negligible, which has no significant impact on the distribution.

We assumed the age distribution to be normal, and calculated its standard deviation ($\sigma_{\rm age}$). Then we use the full width at half maximum (FWHM $=2.355\sigma_{\rm age}$) of this distribution to represent the age spread. The uncertainty of $\sigma_{\rm age}$ is the standard error generated by bootstrap method with 1000 samples under the confidence level of 0.9 using SciPy \citep{2020NatMe..17..261V} plus 5 Myr. We derived a $\sigma_{\rm age}\sim$80.6 Myr for NGC 2423, with an uncertainty of 16 Myr. The derived result might be affected by statistical fluctuations because of the small  sample size (27 stars).

\begin{figure}
    \centering
    \includegraphics[width=0.45\textwidth]{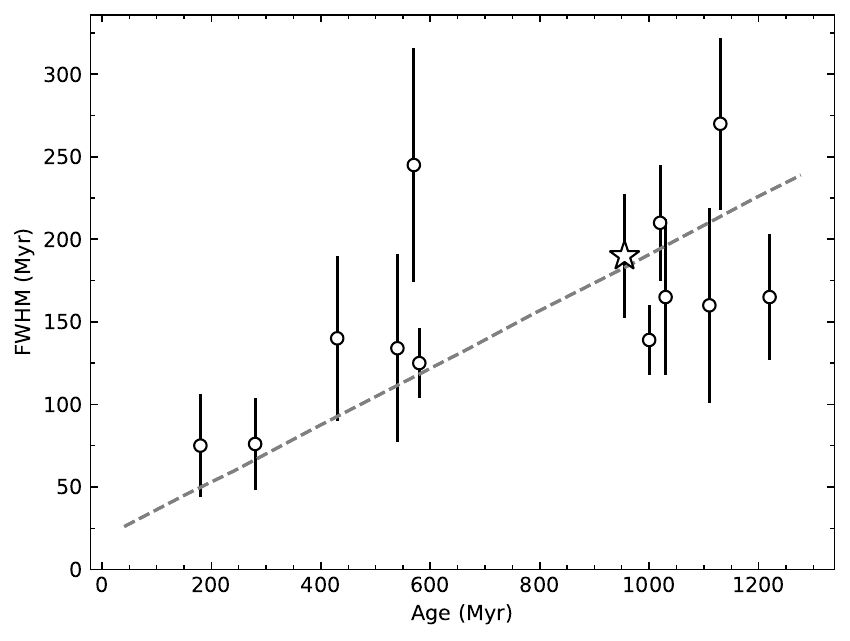}
    \caption{The correlation between the age of clusters and the their age spread, determined from the width of their eMSTO regions. The data is from \cite{2018ApJ...869..139C}, with the empty pentagram representing NGC 2423.
    }
    \label{fig:fwhm2}
\end{figure}

We combined the derived age spread of NGC 2423 and clusters analysed by \cite{2018ApJ...869..139C} (see Figure 11 of their paper for details), shown in Figure \ref{fig:fwhm2}. \cite{2018ApJ...869..139C} found that the eMSTO increases with cluster age when interpreted as a pure age spread. The derived age spread and the cluster age of NGC 2423 fit well with the correlation they found. 

\begin{figure}[t]
    \centering
    \includegraphics[width=0.45\textwidth]{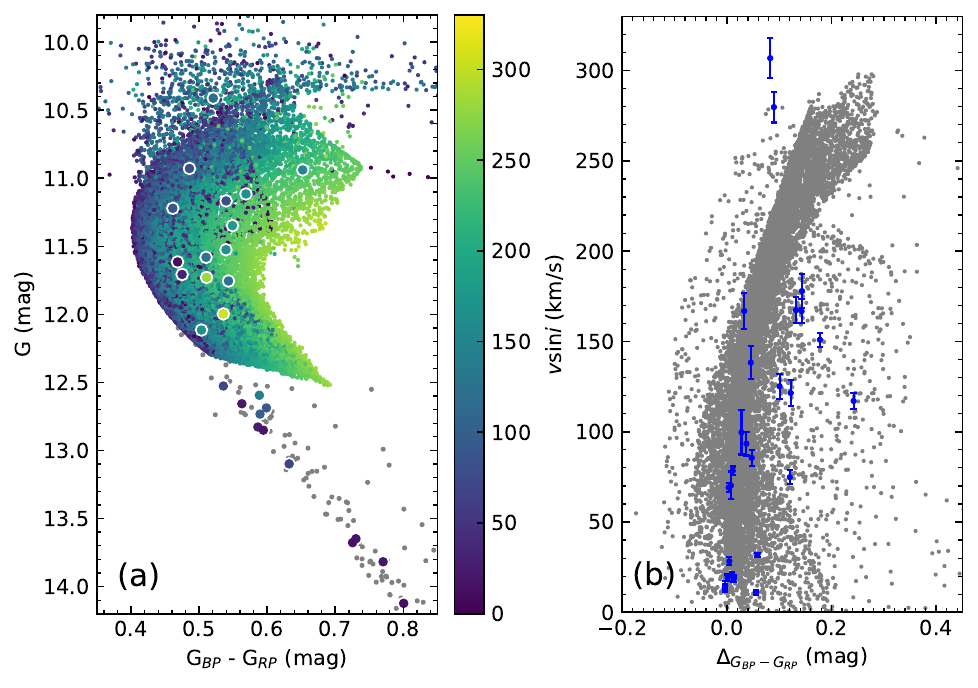}
    \caption{(a) The CMD displays observed stars (represented by large circles) with measured $v\sin{i}$ (color coded), alongside simulated stars (small dots). (b) $\Delta(G_{BP}-G_{RP})$ versus $v\sin{i}$ for observed stars (blue dots with error bars) and simulated stars (grey dots).
    }
    \label{fig:vsini}
\end{figure}

The member stars in MSTO region have a large $v\sin i$ spread (\SI{11.08}{\kilo\meter\per\second}-- \SI{306.90}{\kilo\meter\per\second}), as listed in Table \ref{tab:full_table}. Stars with larger $v\sin i$ are located at redder parts in the MSTO region as shown in Figure \ref{fig:vsini}(a).

To find out whether the stellar rotation is the main cause for the eMSTO in NGC 2423, we use SYCLIST's model \citep{2013A&A...553A..24G, 2014A&A...566A..21G} to generate a coeval simple stellar population with different rotation rates (small colored dots shown in Figure \ref{fig:vsini}(a)). The stellar rotation distribution is generated based on \cite{2010ApJ...722..605H}. The binary fraction after completeness correction of 85 open clusters investigated by \cite{Pang2023AJ....166..110P} is in the range of $\sim$10$\%$ and $\sim$50$\%$. Based on Figure \ref{fig:members}, it appears that the binary fraction of NGC 2423 is relatively low when assessing the unresolved binaries visually. In the simulation, we consider a binary fraction of 30\% with Z=0.014. The synthetic cluster contains 30,000 member stars with mass of \SI{1.7}{M_\odot}$\sim$\SI{15}{M_\odot} \citep{2013A&A...553A..24G}. The gravity darkening law is from \cite{2011A&A...533A..43E}, while the limb darkening effect is not included and the colors$-T_\mathrm{eff}$ calibration from \cite{2011ApJS..193....1W} is implemented. The slow rotators are located in the blue side of the MSTO region and the fast rotators are located in the red side. 

 We plotted a $\Delta$(G$_{BP}-$G$_{RP}$)$-v \sin i$ diagram, in which $\Delta$(G$_{BP}-$G$_{RP}$) is defined as the stellar G$_{BP}-$G$_{RP}$ deviation from that of the best-fitting MIST isochrone in the same Gaia G-band magnitude. The relation between the $v\sin i$ and the color deviation for NGC 2423 and the synthetic cluster is displayed in Figure \ref{fig:vsini}(b). With the exception of the three observed stars exhibiting very high velocities (above  \SI{250}{\kilo\meter\per\second}), all others fall within the range of the synthetic cluster's $\Delta$(G$_{BP}-$G$_{RP}$)$-v\sin i$ distribution.

\subsection{Properties of tidally locked binaries} \label{sec:rv}
Although all the stars in our sample have been observed spectroscopically only once, which limits our ability to identify specific stars with close tidal companions, we can use statistical analysis to predict the presence of a substantial number of tidal binaries within a stellar population. To do this, we compared the RV dispersion of observed members with $v\sin i$ lower than \SI{71}{\kilo\meter\per\second} and a synthetic RV dispersion assuming those members are all unresolved tidally-locked binaries. 

\begin{figure}
    \centering
    \includegraphics[width=0.45\textwidth]{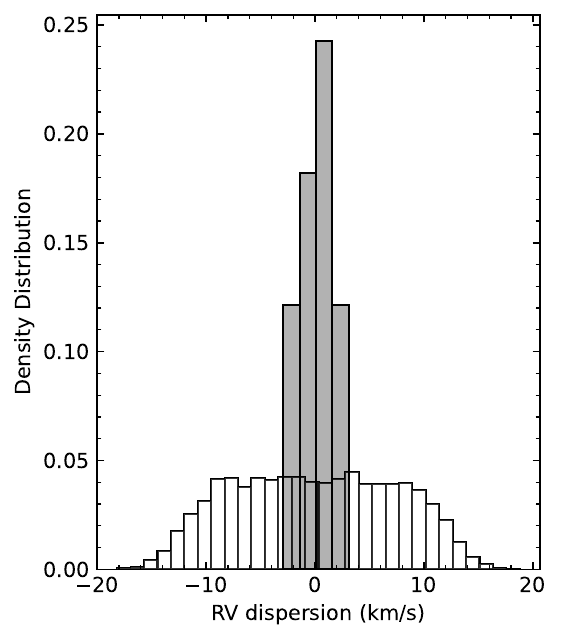}
    \caption{Density distributions of observed NGC 2423 members with $v\sin i <$ \SI{71}{\kilo\meter\per\second} RV dispersion (empty bars), and the synthetic RV dispersion (grey bars).
    }
    \label{fig:RV dispersion}
\end{figure}

Following the expression in \cite{hechenyu_2023MNRAS.525.5880H}, the RV of the tidally-locked component is expressed as 
\begin{equation}
    RV_{\rm obs} = RV_{\rm c} + \frac{a}{R}v\sin i\frac{q}{(1+q)}\cos \phi_t,
\end{equation}
where $RV_{\rm c}$ is the centroid RV; $R$ is the radius of the primary star; $a$ and $q$ refer to the separation and mass ratio between the primary star and its companion, respectively; $\phi_t$ represents the phase of the binary orbit at a given observation time, denoted by $t$.  Based on the equation in \cite{Hurley_2002MNRAS.329..897H}: 
\begin{equation}
    \frac{1}{\tau_{\rm sync}}=5\times2^\frac{5}{3}\left(\frac{GM}{R^3}\right)^\frac{1}{2}\frac{MR^2}{I}q^2\left(1+q\right)^\frac{5}{6}E_2\left(\frac{R}{a}\right)^\frac{17}{2},
\end{equation}
where $G$ refers to the gravitational constant; $M$ represents the mass of the primary star. $I$ is the moment of inertia; $E_2$ is a second-order tidal coefficient which can be calculated by 
\begin{equation}
    E_2 = 1.592\times10^{-9}M^{2.84}
\end{equation}  
\cite{Zahn_1975A&A....41..329Z}.  In this work $q$ is setted to 0.5 (a large $q$ will lead the observed target too red to populate the bMS). $M$ is estimated by the best fitting isochrone. $R$ is estimated by the empirical relation: $R=1.01M^{0.724}$ for 0.1$M_\odot$$<$$M$$<$$18.1M_\odot$ \citep{Demircan_1991Ap&SS.181..313D}. We aim to derive the relation between synchronization time and $a$/$R$. We then generated the synthetic RV distribution with $7\le a/R\le 8$ (corresponding to a synchronization time scale between 1 Gyr -- 10 Gyr). In this case, RV of the binary component of a tidally-locked binary is expressed as
\begin{equation}
    RV_{\rm obs} = RV_{\rm c} + v_{\rm orb}\sin j\cos \phi_t
\end{equation}
\cite{hechenyu_2023MNRAS.525.5880H}, where $v_{\rm orb}$ is the orbital velocity of the primary star and the angle $j$ represents the inclination of the binary orbital rotation axis.  We calculated the orbital periods, $T_{\rm orb}$, utilizing Kepler's Third Law. The $v_{\rm orb}$ were calculated applying $v_{\rm orb}=2\pi qa/(1 + q)T_{\rm orb}$. We applied \SI{19.9}{\kilo\meter\per\second} (the mean RV of observed members with $v\sin i$ lower than \SI{71}{\kilo\meter\per\second}) to $RV_{\rm c}$.  The $\cos \phi_t$ and $\sin j$ is randomly determined in uniform distributions. We proceeded this process for 1000 times and recorded RV variations for each run. The comparison of density distribution of RV dispersion between the observed and the synthetic samples is shown in Figure \ref{fig:RV dispersion}.

\begin{figure}
    \centering
    \includegraphics[width=0.45\textwidth]{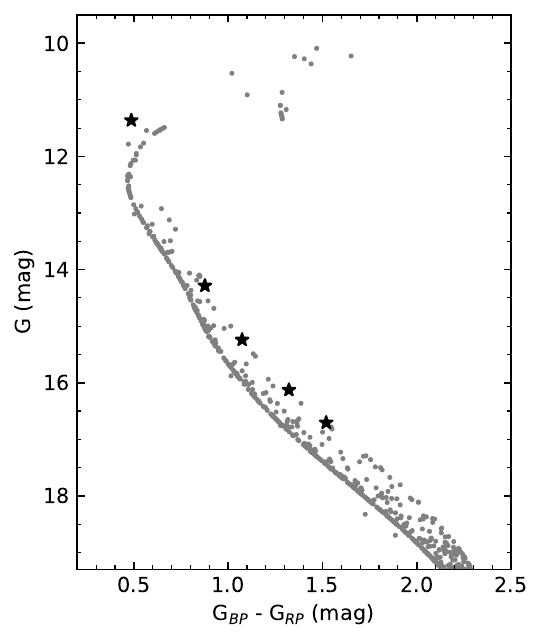}
    \caption{Location of tidally-locked binaries (black  pentagrams) in the CMD of the synthetic {\tt\string PeTar} cluster.
    }
    \label{fig:petar_cmd}
\end{figure}

Furthermore, we identified the tidally-locked binaries (i.e. binaries with the same orbital and rotational period) in the {\tt\string PeTar} synthetic cluster. Their distribution in the CMD is shown in Figure \ref{fig:petar_cmd}. We find that all of the tidally-locked binaries are equal-mass binaries, which means only those ``twins'' can be formed in the age of NGC 2423. Our result is similar to the result from \cite{wangli_2023ApJ...949...53W},

\subsection{Identification of a ``twins''}
\begin{figure}
    \centering
    \includegraphics[width=0.45\textwidth]{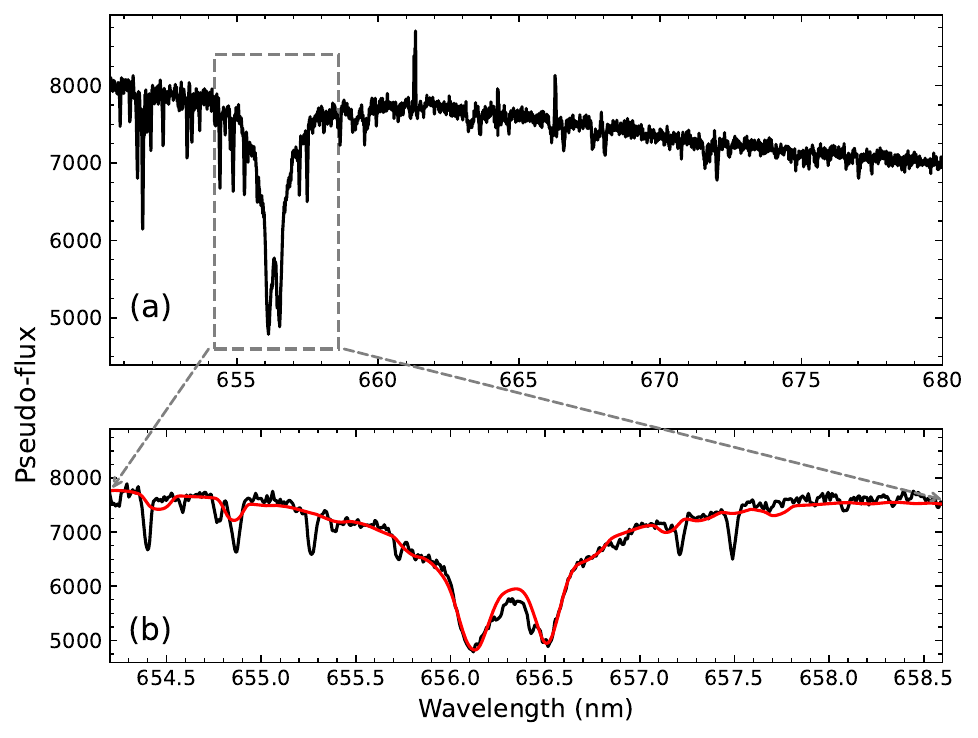}
    \caption{(a) The spectra of the spectroscopic binary stars detected in NGC 2423. (b) A zoomed-in section of panel (a), displaying our fitted model (red curve) for the spectra.
    } 
    \label{fig:td_spec}
\end{figure}

We immediately conducted observational validation of the above results. We searched the ESO archive for spectra of stars near the equal-mass binary loci of NGC 2423. We found that there are only two equal-mass binaries have their spectra accessible from the database. Interestingly, from the limited set of two stellar spectra, we promptly detected a spectroscopic binary star (RA: \SI{114.28}{\degree}, Dec: \SI{-13.89}{\degree}). The binary spectrum (PI: Zaworska Paulina, ID: 098.C-0217) is displayed in Figure \ref{fig:td_spec} and its loci in CMD is presented as the red pentagon in Figure \ref{fig:members}(b). We Applied the approach as mentioned in \ref{sec:vsini} to calculate RVs and $v\sin i$ except combining two sets of theoretical spectra together to fit the binary spectrum. The uncertainties of RVs are derived from the range corresponding to $\chi^2$+1. The RVs of each binary components are $-75.9\pm3.9$\SI{}{\kilo\meter\per\second} and $105.3\pm4.1$\SI{}{\kilo\meter\per\second} (see the fitting in Figure \ref{fig:td_spec}). Given that the mean radial velocity of observed members is \SI{19.9}{\kilo\meter\per\second}, we can estimate the mass-ratio of the binary system of $q\sim0.9$. This finding matches with the location of this binary star on the CMD (in fact, it even appears a bit brighter than the equal-mass binary loci). 

We then derived the light curve of this binary system from the Zwicky Transient Facility(ZTF) Data release 20 \footnote{\url{https://www.ztf.caltech.edu/ztf-public-releases.html}}. This binary system is designated as ZTF J0737-1353. ZTF is a photometric survey conducted using a broad 47 $\textrm{deg}^2$  field-of-view camera installed on the Samuel Oschin 48-inch telescope at the Palomar Observatory \citep{bellm2019, graham2019, dekanyztf, masci_ztf}. We conducted a period search analysis on two sets of time-series data, including the ZTF $g$- and $r$-band data, utilizing the Conditional Entropy algorithm to accurately determine the periodicity of irregularly sampled astronomical time-series data \citep{2013MNRAS.434.2629G}. This method involves minimizing the conditional Shannon entropy while folding the light curve at a trial period, ensuring precise identification of the system's periodicity.

\begin{figure}
    \centering
    \includegraphics[width=0.5\textwidth]{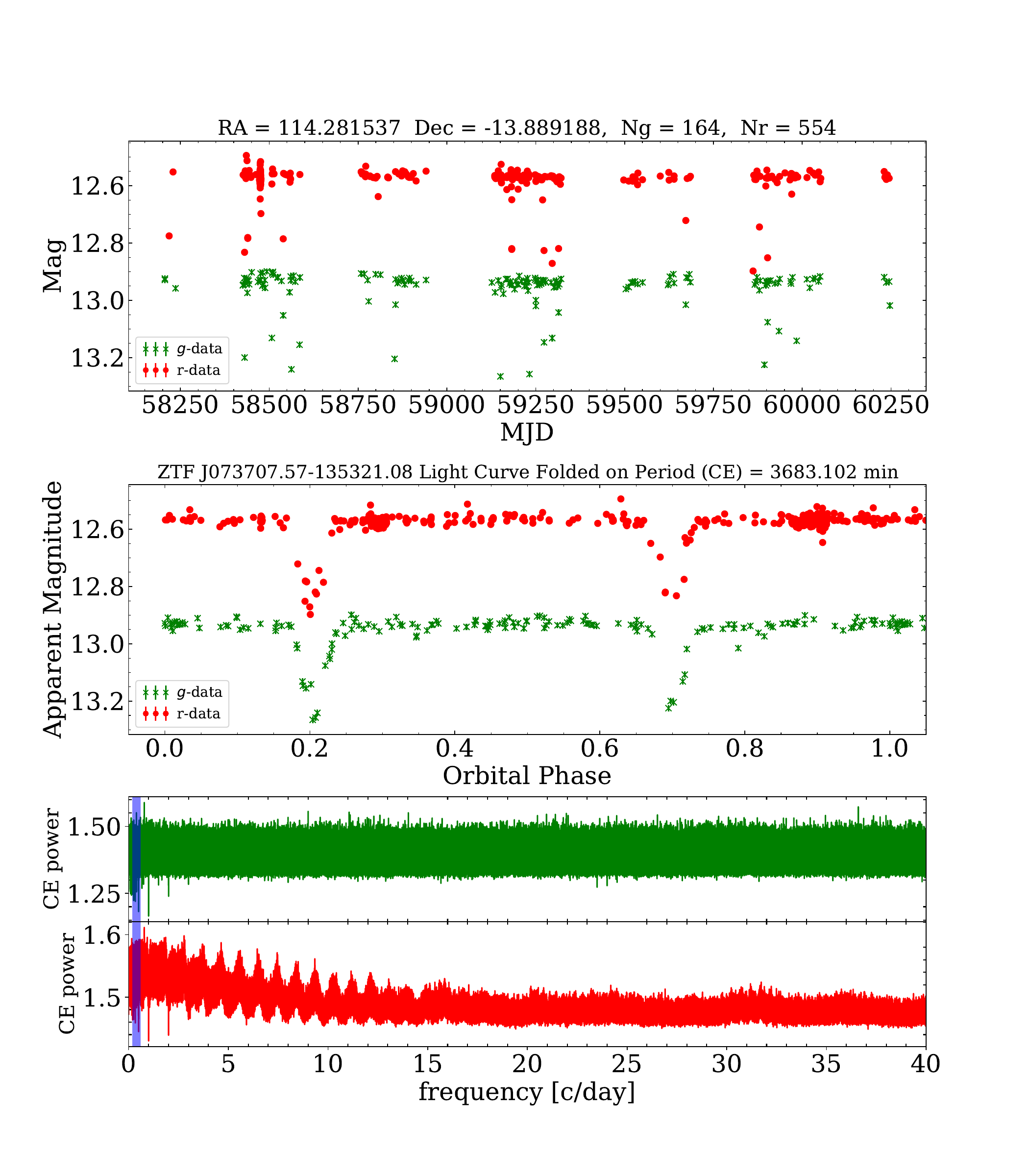}
    \caption{The figure  illustrates the light curves of ZTF J0737-1353, with the top panel displaying the raw ZTF light curves in the $g$-band (green) and $r$-band (red), the middle panels showing the phased ZTF light curves for the $g$-band and $r$-band folded at a period of 3683 minutes, and the bottom panels depicting the frequency spectra of the Conditional Entropy algorithm.
    }
    \label{fig:light_curve}
\end{figure}

The light curve indicates that the binary has an orbital period of \SI{3683.1}{\minute} (i.e. \SI{61.39}{\hour}), as shown in Figure \ref{fig:light_curve}. As expected, the primary and secondary eclipses of this binary star show almost identical depths, with no prolonged flat-bottomed phase at the eclipse minima, indicating its nature as an equal-mass binary system with companion stars of comparable size and brightness. The phase offset between the primary and secondary eclipses is nearly precisely 0.5, implying an orbital eccentricity of $\sim$0 for this binary system, thus the circularization of its orbit. 

To find out whether this binary is tidally-locked, we calculated rotational periods of the binary. First, the $v\sin{i}$ for the primary is $28.9\pm0.9$ \SI{}{\kilo\meter\per\second} and for the secondary companion, is $v\sin{i}=36.1\pm1.9$ \SI{}{\kilo\meter\per\second} (The uncertainties are taken from the range corresponding to $\chi^2$+1). We calculated the mass and size of each component by interpolating the most suitable isochrone, assuming a mass ratio of $q$=$M_2$/$M_1$=0.9. The results are $M_1$=1.419$M_{\odot}$,  $M_2$=1.265$M_{\odot}$ and $R_{\rm 1}$=1.485$R_{\odot}$, $R_{\rm 2}$=1.251$R_{\odot}$. By applying Kepler's laws, we calculated the semi-major axis of the system to be approximately $a=10.93R_{\odot}$. Utilizing the depths of the light curves, we determined the inclination ($i$) of the binary system to be 85.26$^{\circ}$, i.e., the orbital plane of this binary system is nearly parallel to the line of sight. Combining the above calculations, we ultimately obtained the rotation periods of the primary and secondary stars to be 62.47$^{+2.12}_{-1.98}$ \SI{}{\hour} and 41.66$^{+2.29}_{-2.05}$ \SI{}{\hour}, respectively. The primary star is nearly fully tidally locked, with its ratio between the rotation and orbital periods of $P$/$P_{\rm orb}$=1.02$\pm$0.03.

\begin{deluxetable}{c|cc}
\label{tab:binary}
\tablecaption{Fitting parameters of the detected double-line spectroscopic binary}
\tablehead{
    \colhead{} &
    \colhead{Primary} &
    \colhead{Secondary}
}
\startdata
Mass ($M_{\odot}$) & 1.419 & 1.265 \\
Radius ($R_{\odot}$) & 1.485 & 1.251 \\
$T_{\rm eff}$ ($K$) & 6200$\pm200$ & 6600$\pm200$ \\
$\log{g}$ (dex) & $\sim$3.5$-$$\sim$5.0 & 4.1$-$$\sim$5.0 \\
$RV$ (\SI{}{\kilo\meter\per\second}) & 105.3$\pm$4.1 & $-$75.9$\pm$3.9  \\
$v\sin{i}$ (\SI{}{\kilo\meter\per\second}) & 28.9$\pm$0.9 & 36.1$\pm$1.9 \\
$P$ (\SI{}{\hour}) & 62.47$^{+2.12}_{-1.98}$ & 41.66$^{+2.29}_{-2.05}$ \\
$P$/$P_{\rm orb}$ & $1.02\pm0.03$ & 0.68$^{+0.04}_{-0.03}$ \\
\enddata
\tablecomments{The inclination of the binary system is $i$=85.26$^{\circ}$}
\end{deluxetable}

\section{Discussion} \label{sec: discussion}
The first scientific implication provided by our results is that the eMSTO in NGC 2423 is driven by different stellar rotations rather than an age spread. It is not surprising. Firstly, age spread cannot explain why NGC 2423 precisely matches the linear correlation between the age spread obtained in \cite{2018ApJ...869..139C} and the cluster age. (i.e., why the age spread in NGC 2423 is larger than in younger clusters but smaller than in older clusters.) The current stellar formation scenario cannot naturally account for this result. On the other hand, stellar rotation can better explain this phenomenon, as the impact of rotation on the broadening of the MS is limited. The younger the cluster, the more sensitive it is to age spread. Therefore, the eMSTO produced by rotation in younger clusters may appear to stem from a smaller age spread. Indeed, the positive correlation between age spreads and clusters' ages has been demonstrated in the model proposed by \cite{2015MNRAS.453.2070N}, assuming that the eMSTO phenomenon is primarily influenced by stellar rotation.

\begin{figure}
    \centering
    \includegraphics[width=0.45\textwidth]{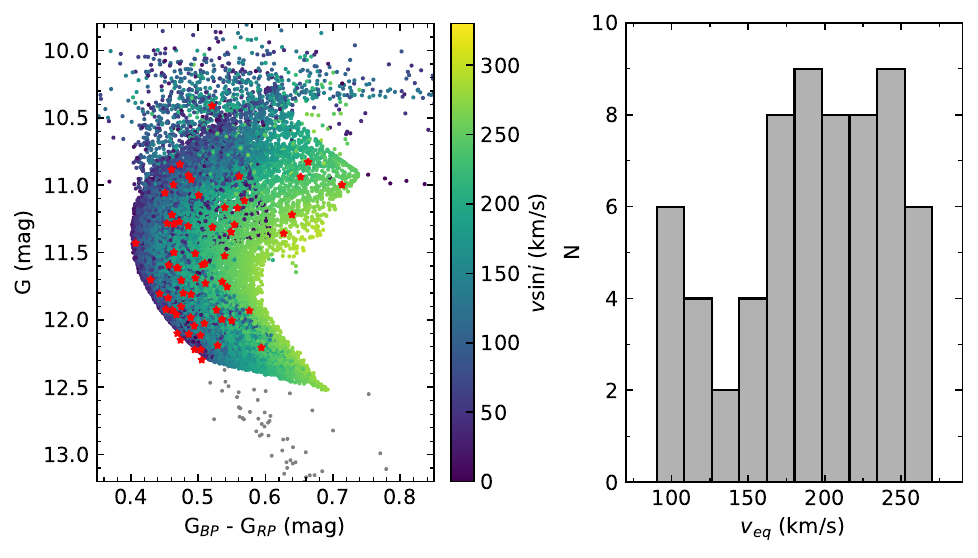}
    \caption{The distribution (panel (b)) of the estimated equatorial rotational velocities ($v_{eq}$) for selected stars in the MSTO region (red stars in panel (a)).
    }
    \label{fig:v_eq_distribution}
\end{figure}

Additionally, our analysis of stellar $v\sin{i}$ measurements also exhibits a consistent positive correlation between $v\sin{i}$ and color, indicating that stars with higher rotation velocities tend to exhibit redder on average. This result is similar to \cite{2019ApJ...876..113S} and \cite{Kamann2020MNRAS.492.2177K}, who found that stars in the redder part of the eMSTO show larger rotation rates in the intermediate-age Galactic cluster NGC 5822 and LMC cluster NGC 1846, respectively. This is precisely the expected outcome of the gravity darkening effect induced by rotation (Figure \ref{fig:vsini}(b)). Owing to constraints of our sample size, we cannot determine the presence of a younger bMS population, as reported  in \cite{D'Antona_2017NatAs...1E.186D} for other LMC clusters. However, we can employ statistical methods to examine whether NGC 2423 exhibits two relatively separated populations with different rotations, similar to \cite{D'Antona_2017NatAs...1E.186D,2022ApJ...938...42H,2019ApJ...876..113S}. We used the same method as \cite{2019ApJ...876..113S} to estimate the equatorial rotation rate ($v_{eq}$) of 64 available eMSTO stars of NGC 2423. The $v_{eq}$ of the stars were estimated as the average $v_{eq}$ of 50 synthetic stars closet to them in the CMD, generated from the SYCLIST model and could properly reproduce the $v\sin i$ and color distributions of the eMSTO stars (see Figure \ref{fig:v_eq_distribution}(a)). Figure \ref{fig:v_eq_distribution}(b) shows the distribution of the estimated $v_{eq}$. The eMSTO stars seem to be separated into one slowly rotating population with $v_{eq}$ peaking at \SI{100}{\kilo\meter\per\second} and one rapidly rotating population with average $v_{eq}$ around \SI{200}{\kilo\meter\per\second}. This result is consistent with the finding of \cite{Kamann2020MNRAS.492.2177K}, who detected a bimodal $v\sin i$ distribution among the eMSTO stars of NGC 1846. As the mass range of the eMSTO stars of NGC 2423 is comparable to the mass range for the appearance of split MS \citep[$M>1.6 M_{\odot}$;][]{2018MNRAS.477.2640M}, the detected double-peaked $v_{eq}$ distribution may be the continuity of a bimodal rotation rate distribution during their MS time, indicating that NGC 2423 may exhibit separated MS during its younger age. But we emphasize that the small number statistics may influence the obtained results inferred from the $v_{eq}$ distribution. 

Assuming that the rotation distribution in NGC 2423 reflects the bimodal rotation distribution observed in LMC clusters, following the hypothesis proposed by \cite{D'Antona_2017NatAs...1E.186D}, bMS stars are expected to have undergone rotational deceleration. Our results indicate that the deceleration is unlikely to be attributed to tidal interactions in binary systems. The absence of significant radial velocity dispersion among the slowly rotating stars in NGC 2423 allows us to statistically exclude the possibility of a substantial portion of them being part of tidal binary systems, This is also similar to the findings of \cite{hechenyu_2023MNRAS.525.5880H} for cluster NGC 2422. Our numerical simulations agree with the results of \cite{wangli_2023ApJ...949...53W}, even though the star clusters they studied are much younger than ours. We found that, even after $\sim$1 Gyr, only binary systems with similar masses ($q\sim$1) are likely to undergo tidal locking among the MS stage. This is because the tidal fields of low-mass ratio binaries are weak on average, unless their initial semi-major axis distribution is biased towards close proximity, they cannot reach tidal locking within $\sim$1 Gyr.

The discovery of tidally locked binary stars in NGC 2423 closely matches out simulation. It is important to highlight that this binary system still has some unresolved problems. For example, why does it appear obviously redder than the equal-mass binary loci in the CMD? Why is the primary star tidally-locked while the secondary star is not fully tidally-locked yet? These queries will be further explorations in our forthcoming research. Here we speculate that possible reasons could include errors, such as the variability and point spread functions residuals of this binary system potentially causing larger photometric uncertainties than expected, or the presence of biases in traditional spectral fitting methods for double-lined spectroscopic binaries, among other factors. It is also possible that these issues are due to real physical reasons. For example, this binary may have at least one of its component more evolved than normal MS stars, or it is affected by greater local extinction. The primary star is more affected by tidal field than the secondary star due to their different structures: the primary star may has a radiative envelop while the secondary star has a convective envelope. 

Since we only have two spectra of stars close to the equal-mass binary loci, our discovery could be a coincidence, or it could indicate the validity of our model (especially given the lack of clear evidence for tidal binaries among the 11 slowly rotating stars). If the latter, it would implicates that at least within open clusters younger than $\sim$1 Gyr, all tidally-locked binaries must be also ``twins'' systems, i.e., binaries with two identical components. This is significant as it may suggest that the dynamical interactions within the star cluster play a crucial role in the formation of twins. It also implies that many exotic objects, such as double white dwarfs, can effectively form through the dynamical pathways within the star cluster, providing valuable guidance for the next generation of space-based gravitational wave detectors in the search for stellar gravitational wave electromagnetic counterparts \citep[e.g.,][]{2023ApJS..264...39R}.

\section{Conclusion} \label{sec: conclusion}

In this work, we investigated NGC 2423, a Galactic open cluster with age of $\sim$ 1 Gyr. The cluster exhibits an eMSTO. We aim to study its properties related to stellar rotation and binarities, to constrain their role on the formation of the eMSTO. Our main conclusions can be summarized as below, 

\begin{itemize}
\item[1.] The eMSTO in NGC 2423 is attributed to variations in stellar rotation rates rather than age differences. The equatorial rotational velocity exhibit a possible bimodal distribution. 

\item[2.] The slowly rotating stars in NGC 2423 show a lack of tidal binaries. If these slowly rotating stars were once rapidly rotating, then tidal locking is not the major deceleration mechanism.

\item[3.] Our numerical simulations indicate that in NGC 2423 and similar-aged open clusters, tidally locked binaries are likely to be equal-mass binaries. In cases where they cannot be resolved by telescopes, they are unlikely to form the blue portion of the MS in CMDs.

\item[4.] In one of the two equal-mass binaries we observed in NGC 2423, we found evidence of tidal locking. If this is not merely a coincidence, it could partially validate the accuracy of our model. Searching for such twins binaries in more star clusters may hold theoretical significance.
\end{itemize}


\begin{acknowledgments}

This work was supported by the National Natural Science
Foundation of China (NSFC) through grant 12233013 and 12073090. This work has made use of data from the European Space
Agency (ESA) mission {\it Gaia}
(\url{https://www.cosmos.esa.int/gaia}), processed by the {\it Gaia}
Data Processing and Analysis Consortium (DPAC,
\url{https://www.cosmos.esa.int/web/gaia/dpac/consortium}). Funding
for the DPAC has been provided by national institutions, in particular
the institutions participating in the {\it Gaia} Multilateral
Agreement. This research has used the POLLUX database
(\url{http://pollux.oreme.org}), operated at LUPM (Universit\'e
Montpellier--CNRS, France), with the support of the PNPS and
INSU.
We thank Giacomo Cordoni (University of Padova) for providing the data of the clusters analysed in \cite{2018ApJ...869..139C}.

\end{acknowledgments}

\software{{\tt\string StarGO} \citep{Yuan_2018}, {\tt\string PeTar} \citep{wanglong_2020MNRAS.497..536W}, MIST \citep{2016ApJ...823..102C}, Astropy \citep{2013A&A...553A..24G,Price-Whelan_2018}, Matplotlib \citep{2007CSE.....9...90H}, SciPy \citep{2020NatMe..17..261V}, PyAstronomy \citep{2019ascl.soft06010C}, Astrolib PySynphot \citep{2013ascl.soft03023S}, SYCLIST \citep{2013A&A...553A..24G, 2014A&A...566A..21G}, TOPCAT \citep{2005ASPC..347...29T}.
}

%





\bibliography{sample631}{}
\bibliographystyle{aasjournal}



\end{CJK*}
\end{document}